\documentstyle[12pt,prl,aps]{revtex}
\newcommand{\pub}[4]{{\em #1 }{\bf #2}, #3 (#4)}

\newcommand{\jpa}{J. Phys. A}

\newcommand \be  {\begin{equation}}
\newcommand \bea {\begin{eqnarray} \nonumber }
\newcommand \ee  {\end{equation}}
\newcommand \eea {\end{eqnarray}}
 \def\(({\left(}
 \def\)){\right)}
 \def\[[{\left[}
 \def\]]{\right]}
\def\nn{\nonumber}
\def\bi{\bibitem}

\def\o{ \omega}

\def\eps{\epsilon}
\def\cP{\cal {P}}

\begin{document}

\title{
Random systems and replica field theory}
\author{ Marc M\'{e}zard$^{(1)}$}
\address{
CNRS-Laboratoire de Physique Th\'{e}orique de l'ENS$^{(1)}$\\
( $^{(1)}$Unit\'e propre du CNRS, associ\'ee \`a l'ENS et \`a
l'Universit\'e de Paris XI)\\
24, rue Lhomond; F-75231 Paris Cedex 05; France\\
e-mail: mezard@physique.ens.fr}
\date{February 1995}
\maketitle
\begin{abstract}
Summary of lectures given at the 1994 Les Houches Summer school.

{\it Preprint LPTENS 95/10}
\end{abstract}

\baselineskip 18pt plus 2pt minus 1pt
\section{Introduction}

There are two aspects to the statistical physics of systems
with quenched disorder. One is the static aspect, namely to understand
the properties of the Boltzmann Gibbs distribution, the other is
the dynamical problem. In these lectures I would like to concentrate
on some recent developments about the statics. The interested reader
can find some brief description of the recent works on dynamics
in \cite{icmp94}.
A system with quenched disorder contains two types of variables,
$q$ and $t$, which equilibrate on very different time scales.
The thermalized variables $t$ are assumed to reach thermal equilibrium
(this is an assumption of the static problem, it is
in general  not
true for real materials, and then the dynamical approach is needed
at low temperatures). The quenched variables $q$ are given a priori
and do not thermalize. In the problems we shall study there are
many quenched variables, which means a number which diverges in the
thermodynamic limit where the number of quenched variables diverges.
One sample of the disordered system corresponds to one set of values
of $q$. We would need an infinite amount of information to just
describe one sample. So in practice what we know is the probability
distribution, $\cP[q]$, of $q$.

In practice there are many systems in this category which have been
studied in recent years. A canonical example is the spin glass problem
\cite{Anderson,mpv}. The techniques and ideas from the spin glass
 mean field theory
have been applied to many other systems, like directed polymers
(see for
instance \cite{KarZ,DerSp,parisi90,MM90}). One could also mention the Random
Field Ising Model (RFIM) \cite{nvby,MezY}, the protein
folding problem
\cite{GarOrl,ShaGut}, and several problems outside of the usual physics
ones, i.e. where the 'energy' function is not a real physical
energy, like neural networks (see for instance \cite{amit}),
or optimization problems
(see for instance \cite{mpv}).

Thermal equilibrium means that the probability of a given configuration
of thermalized variables is given in terms of the
energy $H[q,t]$ by the Boltzmann weight:
\be
P_q[t]={1 \over Z[q]} \sum_{[t]} e^{-\beta H[q,t]}
\ee
where $Z[q]$ is the partition function. So there is one such
probability distribution for each sample. We face a difficult problem
which is to characterize the set of the $P_q[t]$. Fortunately it
turns out that some quantities are self averaging. This
means that they become sample independent (i.e. independent
of $q$) in the thermodynamic limit, a typical central limit
theorem. This is the case for thermodynamic
 quantities, and it is easy to derive this result for finite dimensional
systems with finite range and bounded interactions. The properties
which are not self averaging may also be interesting (for instance
the nature of the ground state of the travelling salesman problem,
or the folding of a protein), but this
is an algorithmic problem. Here we shall keep to analytic
studies of self averaging
quantities and properties.

One standard approach of statistical physics uses mean field methods
to work out the phase diagram of the system. This is efficient
away from the second order phase transitions. Around such transitions
  the renormalisation
group can be used. It has been gradually understood that in
many disordered systems this very general and powerful approach
fails. The fundamental reason for this failure is that the
free energy functional possesses many secondary minima (metastable
states) which can be quite different from the minimum, and are
not taken into account in the standard approach. A striking
example is provided by the RFIM. It has been shown that, order
by
order in perturbation theory the critical
exponents of a $d$ dimensional Ising system with a quenched
random field are equal to those of the pure system in dimension
$d-2$ \cite{py}. This is one of the few results in field theory
which is known to hold to all orders in perturbation, and a
beautiful derivation can be obtained through supersymmetry
\cite{ps79}. However this result
differs from the simple domain wall
argument of Imry and Ma \cite{im75}, and it would predict a lower
critical dimension equal to 1, while there are by now exact
results which show that this lower critical dimension is equal to
two \cite{bk87} \cite{aizenman}.
It has been realized long ago that the failure of the perturbative approach
is related to the existence of many metastable states in this problem
\cite{par82}.

What one needs is a systematic and non perturbative method which
is able to handle the field theory of
 disordered systems with many metastable states. Such a method
can be obtained with the use of replicas, to handle randomness,
together with a gaussian variational method. It turns out that the
possibility of breaking replica symmetry allows to handle the
metastable states. Such an approach was used on the problem
of heteropolymers or proteins \cite{GarOrl,ShaGut}, but this is
a complicated problem which requires additional assumptions
\cite{fmp}. It has been developed as a systematic field theoretic
method to handle manifolds in random media in \cite{MP,MP2},
and used since on several problems. Before turning to this
case, let us just comment on the general meaning of
replica symmetry breaking (rsb). The replica method allows
to compute extensive thermodynamic quantities like the
free energy. Using the property of self averageness, we
need to compute the average of the logarithm of the partition
function, which is written as:
\be
\overline{\ln(Z)}\equiv \sum_{[q]} \cP(q) \ln(Z(q)) = \lim_{n \to 0}
  {\overline{Z(q)^n} -1 \over n}
\ee
The $n$'th power of the partition function can in turn be written
as the partition function for $n$ replicas of the original
sample (with the same disorder). Disorder averaging leads
to  a pure problem (no disorder), with an
 effective attraction between the replicas:
$$
\overline{Z(q)^n}=\sum_{[q]} \cP(q) {Z(q)^n}=
$$
\be
=\sum_{[q]} \cP(q)
 \sum_{t_1,...,t_n}
{\exp{-\beta \(( H[q,t_1]+...+H[q,t_n]\))}}
\equiv
\sum_{t_1,...,t_n}
\exp{-\beta \(( H_{eff}[t_1,...,t_n]\))}
\ee
The replica Hamiltonian $H_{eff}[t_1,...,t_n]$ is obviously
invariant under permutations of the $n$ replicas. The problem
is to know whether this symmetry is preserved or spontaneously
broken in the thermodynamic limit. As first found in spin
glass theory (for a review see \cite{mpv}), such a spontaneous
symmetry breaking can occur, and is related to the appearance
of a spin glass phase. Roughly speaking, if the system can have
generically
many metastable states, each replica can condense in one or the
other such state, leading to this rsb phenomenon. The problem is
in the term 'generically': In order for $\overline{Z(q)^n}$ to
be typical of the generic behaviour, we need that $n$ go to zero,
and therefore this rsb requires understanding the properties of
the permutation group when $n \to 0$. This was  achieved by Parisi
in 1979 and is reviewed in \cite{mpv}. A more rigorous
definition of rsb can be found using two real replicas
(i.e. $n=2$), coupled through an extensive energy
term of strength $\eps$  \cite{2rep}. We consider two replicas with the same
disorder
governed by the Hamiltonian:
\be
H_\eps[q,t_1,t_2]=H[q,t_1]+H[q,t_2]-\eps \delta[t_1-t_2]
\ee
(Here the last term is written in a symbolic way. It will in fact
depend on the problem at hand. What I mean is an
extensive term which is the
integral of a local interaction between the two replicas, and the
interaction
is attractive if $\eps>0$ and repulsive otherwise). Calling $F(\eps)$
the quenched average of the free energy for this two replica system,
the onset of rsb is signaled by a non analyticity of $F(\eps)$
at $\eps=0$. The idea is the following: the difficulty in random
systems is that we do not know what is the conjugate field which selects
one given state (such a knowledge depends on the
sample and requires an infinite
 amount of information). In the above procedure each replica plays
the role of the conjugate field for the other one. Practical implementations
of this idea can be found in \cite{2rep} for spin glasses and in
\cite{MM90} for directed polymers.

\section{Manifolds in random media}
Consider a  $D$ dimensional manifold described in the solid on solid
approximation
 by a N component
  vector field
 $\o ( x ) $
where   $x$ is the D
 dimensional vector of internal coordinates.
 The
 Hamiltonian describing the system is the sum of a rigidity term and an
  external potential:
\be
h[ \omega ]={1 \over 2} \int dx \ \sum_{\mu=1}^D \left({\partial \o
   \over \partial x_\mu}\right) ^2
  + \int dx \ V(x,\o(x)).
\ee
The external potential is random and gaussian. Its correlations are given by
\cite{HH}:
\be
 \overline{ V(x,\o) V(x',\o ')} = - \delta^{(D)}(x-x') \  N
  \ f\left({[\o-\o ']^2 \
 \over N}\right).
\ee
 In most of the cases the relevant part of the correlation
  function of the
 potential is its asymptotic behaviour for large transverse distances,
  which we suppose to be
 described by a power law:
     \be
 f\(( {\o^2 \over N} \)) \sim_{\o^2 \gg 1}  \ {g \over 2(1-\gamma)}
 \ \(( {
 \o^2 \over N} \)) ^{(1-\gamma)}.
 \ee
 One rather general class of problems (for a review see for instance
\cite{natruj,MP}) consists in understanding the
transverse fluctuations of the manifold, governed by the exponent
$\zeta$ defined by:
\be
 \overline{<[\o(x+l)-\o(x)]^2>} \ \sim_{l \gg a} \ l^{2
 \zeta}.
\ee
The general model is mainly described by three parameters:
  $D $ is the internal
 dimension of the manifold, $ N=d-D $ is its codimension, and  $\gamma$
 characterizes the large
 distance correlations of the potential.
For $N=1$ the manifold is an interface, for $D=1$ it is a directed polymer,
for $N=D$ it describes the elastic deformations of a $D$ dimensional
crystal with impurities, and for $D=3,N=2$ it describes the elastic
deformations of a vortex lattice in certain pinning regimes of type
two superconductors \cite{boumezyed}.

\section{Thermal fluctuations without disorder}
If $V=0$, the fluctuations are purely thermal. The propagator is
$1/k^2$, leading to $\zeta=0$ if
$D>2$ (the manifold is flat), and $\zeta=(2-D)/2$ if $D<2$
(the manifold is rough), and all these results are independent from $N$.

\section{Random forces}
Another simple case is that of random forces, where the pinning term
of the Hamiltonian is $-\int dx \ f(x) \o(x)$, with
gaussian random forces of strength F. This may be relevant for
the deformation of a cristal lattice on small enough length scales
so that the displacement of an atom is much smaller
than the
correlation length of the potential. The problem is simple
because there are no metastable states. One gets immediately
\be
<\o(k)>={f(k) \over k^2}
\ee
so that:
\be
\overline{<\o^\alpha(k)><\o^\beta(k')>}=\delta^{\alpha\beta}F^2 \delta(k+k')
{1 \over k^4}
\ .
\ee
Clearly the $1/k^4$ term leads to a $\zeta$ exponent which is zero above $D=4$,
and equal to $(4-D)/2$ in $2<D<4$ (In $D<2$ the lateral fluctuations
between two points at distance $x$ depend on the total size of the manifold).
The thermal fluctuations are much weaker since the connected
correlation function is:
\be
\overline{<\((\o^\alpha(k)-<\o^\alpha(k)>\))^2>} ={T \over k^2}
\ .
\ee
The lesson from this computation is that the disorder fluctuations are
dominant, and they lead to a typical "dimensional reduction" result, namely
the fact that the value of $\zeta$ for the disordered system in
$D$ dimension is equal to the value of $\zeta$ for the thermal fluctuations of
the pure system in dimension $D-2$. This is exact for random forces.

\section{Random potential: variational approach}
This is a complicated problem with many metastable states.
A simple argument is that of Imry and Ma
\cite{im75}.
Let me just do it in the case of an interface ($N=1$) in the  RFIM, which
corresponds  to $\gamma=1/2$. A bump of size $\o \simeq w$ on a length scale
$L$
costs an elastic energy $L^{D-2} w^2$, while the typical gain from the
pinning energy is  of order $\sqrt{L^D w}$. Optimizing $w$ leads to
$w \simeq L^{(4-D)/3}$, so that $\zeta=(4-D)/3$. Generalizing this argument
to arbitrary  correlations of the potential, we find:
\be
\zeta={4-D \over 2(1+\gamma)} \ .
\ee
This is a reasonable guess for the exponent $\zeta$, but it is difficult
 to improve it and it has no reason to be exact (a well known counter
example is the case $N=D=1,\gamma=3/2$, where it is known
\cite{HHF,ka87} that
$\zeta=2/3$, while the above formula leads to $\zeta=3/5$).

 What one would like
is a general field theoretic approach similar to that of the previous sections
to handle this problem.
A first step is to use perturbation theory. Although this is not
strictly necessary, it is convenient to introduce replicas. The
effective replica Hamiltonian after disorder averaging is:
\be
H_n = {1 \over 2} \int dx \ \sum_{a=1}^n \ \sum_{\mu=1}^D \left(
 {\partial \o_a\over \partial x_\mu}\right)^2 + {\mu \over 2}
 \int dx \ \sum_{a=1}^n
 \left( \o_a(x) \right)^2
\nn
\ee
\be
+ {\beta \over 2} \int dx \sum_{a,b}
 N f\left( {[\o_a(x)-\o_b(x)]^2 \over N} \right).
\ee
The usual perturbative treatement of the Hamiltonian $H_n$
  consists
 in expanding $f$ in
 powers
 of $ \o$; the quadratic part gives the free propagator $F_{ab} $:
     \be
F_{ab}(k) = {\delta_{ab} \over k^2+n2\beta f'(0)} \ + \ {2\beta f'(0)
 \over k^2(k^2+
 n2\beta f'(0))}.
 \ee
 For  $n \to 0$ , the presence of the  $1/k^4$  term immediately
 leads to $\zeta = (4-D)/2$, as in the random force case.
This result holds at higher orders in perturbation. We face a typical
situation where perturbation theory, which is unable to handle
the metastable states, gives qualitatively wrong results.

The method developed in \cite{MP} uses a variational method
which evaluates the best quadratic Hamiltonian to compute
the quenched free energy with the replica method. I shall describe
only the main ideas here, leaving all the (sometimes complicated)
technical steps aside. The reader is referred to \cite{MP}
for a full presentation. The case $D=0$, studied in
\cite{MPtoy,Villain,engel94} is a somewhat simpler
exercise which is also of interest.
The most general quadratic Hamiltonian is:
\be
H_v = {1 \over 2} \int dk \ G^{-1}_{a,b}(k) \ \o_a(x) \o_b(x)
\ee
Optimizing the variational free energy leads to a simple gap
equation for the self energy  $\sigma_{ab}$:
\be
G^{-1}_{a,b}(k)= k^2 \delta_{ab} -\sigma_{ab}
\ee
\be
 \sigma_{ab} = 2 \beta \hat f' \left({1 \over \beta} \int dk [G_{aa}(k)+
 G_{bb}(k)-2 G_{ab}(k)]\right), \ \ a \ne b
\ee
\be
 \sigma_{aa} = -\sum_{b(\ne a)} \sigma_{ab}.
 \ee
It is easy to seek a replica symmetric solution to these equations,
i.e. to assume that $\sigma_{ab}=\sigma$. One immediately finds again
the famous $1/k^4$ term in the propagator, leading to $\zeta=(4-D)/2$.
This result is wrong mathematically because this saddle point
is not a local minimum of the variational free energy. It is also wrong
physically. The reason is the following: The most general
replica symmetric quadratic Hamiltonian describes a situation
in which:
\be
\overline{Z^n} \simeq \int d\o_a \exp\((-{\beta \over 2} \int dk
\(( k^2 \sum_a \o_a(k) \o_a(-k) -\sigma \sum_{a,b} \o_a(k) \o_b(-k) \)) \))\ .
\ee
Disentangling the square, one gets:
\be
\overline{Z^n} \simeq \int d[f]
\exp\(( -{1 \over 2 \beta \sigma} \int dk |f(k)|^2 \))
\[[
\int  d\o\exp\((-{\beta \over 2} \int dk
\(( k^2  |\o(k)|^2 - \o(k) f(-k)\)) \)) \]]^n \ .
\ee
In this form it is clear that the situation we describe is nothing but a random
force problem, with a quenched random force with a distribution
$\exp\(( -{1 \over 2 \beta \sigma} \int dk |f(k)|^2 \))$. The replica symmetric
quadratic Hamiltonian is bound to describe the random force
problem, a problem without metastable states. To get better results
one needs to break replica symmetry.

The rsb solution has been worked out using the Ansatz developed by
Parisi for the spin glass problem (see the review in \cite{mpv}). This
Ansatz describes hierarchical $n * n$ matrices in the $n\to 0$ limit
by continuous functions on the interval $u \in [0,1]$. Therefore
the self energy becomes a function of this internal parameter
$u$. We get a spectrum of masses, which results in a non trivial
behaviour of the correlation function. The wandering exponent
turns out to be identical to the one derived above
through the Imry Ma argument. Therefore
it provides a microscopic "derivation" of this
result. I shall not reproduce here
these technical computations, but rather sketch in the next
section the physical interpretation of the result. Let me
point out a few properties of the gaussian rsb variational
method. It is a very versatile method which can be used
on many problems. It is basically a Hartree approximation
which becomes exact in the large $N$ limit. A systematic expansion
around $N=\infty$ is thus in principle possible, although technically
difficult. Such an expansion was carried to first order on the RFIM, because
the leading Hartree approximation had no rsb (see \cite{MezY}).
In the case of the random manifold, some attempts have been made
\cite{MPtoy,goldsch}, but the full expansion has not yet been
worked out. Only recently has the replica technical apparutus needed
for this expansion been developed \cite{DKT}.

\section{Physical interpretation of the solution}
It is very instructive to work out the physical content of the
rsb solution, in the same spirit as we did for the rs solution before.
The idea is to consider seriously the quadratic Hamiltonian
with rsb, and deduce what kind of physical situation it
describes. This is not an easy task. Again I will mention
only the results, referring the reader to \cite{MP} for
the derivations. Let us concentrate on one
degree of freedom $\o = \o(x) -\o(0)$.

We have already
seen before that in the rs case $\o$ feels a random force.
It means that for a given sample there is one value
of the force $f$, and the probability distribution of
$\o$ is $c^t \ \exp\((-a (\o-f)^2\))$.

The next stage of approximation
uses a single breaking step in Parisi's hierarchical construction.
Its physical interpretation is as follows:
For each sample one generates a set of favoured values of $\o$,
called $\o_{\alpha}$, together with a set of weights $W_{\alpha}$.
These variables $\o_{\alpha}$ and $W_{\alpha}$ are quenched random variables.
The favoured values $\o_{\alpha}$ are independent variables, with a
distribution
\be
 {\cal P} (\o_\alpha) \ = \ {1 \over {2\pi q_1}^{N/2}} \ \exp \((
 {-{\o_\alpha^2 \over 2q_1}} \))
\ee
As for the weights $W_{\alpha}$, they are derived from some
 "free energy"
  variables   $f_{\alpha}$ through:
\be
W_{\alpha} = {e^{-\beta f_{\alpha}} \over \sum_{\nu} e^{-\beta
f_{\nu}}}.
 \ee
The $f_{\alpha}$ are independent
 random variables with an exponential distribution such that
 the average number of states with free energy  less than $f$ (i.e. weight
 $W_{\alpha}$ greater than $e^{-\beta f}$) is:
\be
 {\cal N}(f) \sim e^{\rho f},
 \ee
For a given sample, that is given the variables $\o_{\alpha}$ and $W_{\alpha}$,
the Boltzmann probability for $\o$ is given by:
\be
P(\o)= c^t \ \sum_\alpha W_{\alpha} \exp \((-{[\o-\o_\alpha]^2 \over 2q_0}\))
\label{P1rsb}
\ee
This distribution of $\o$ is characterized by three
numbers $q_0,q_1,\rho$, which are the variational parameters which are
determined by the variational method at this one step rsb stage.
The above formula (\ref{P1rsb}) is somewhat surprising. It parametrizes
the Boltzmann distribution of the degree of freedom $\o$ by
a weighted sum of gaussians. Such a representation clearly allows for
the existence of several metastable states contributing simultaneously
to $P(\o)$, which is nice. The surprising point is that such a
situation is possible using a  gaussian approximation. The power
of the method comes form the combined used of the gaussian variational
method together with the breaking of replica symmetry.  An elementary
example would be to look at a particle in a
symmetric double well potential,
and approximate its Boltzmann weight by a gaussian. At low
 enough temperature
the best gaussian is shifted from the origin and breaks the inversion
symmetry in the problem. To restore it one approximates the Boltzmann
distribution
of the particle by the sum of the two symmetric
shifted gaussians, and one  finds
a pretty good approximation to the exact distribution at low temperatures.
The situation is exactly analogous here, but the symmetry at hand is
the replica permutation symmetry.

The full construction consists in iterating the above procedure. For
two steps of rsb one must divide each gaussian in (\ref{P1rsb}) into
a weighted sum of subgaussians, introducing two new parameters (one
'q' parameter for the width of the subgaussians, one '$\rho$'
parameter for the distribution of their relative weights). The
construction is then iterated an infinite number of times, adding each
time two new variational parameters. Clearly in this way we can
generate very complicated probability distributions for $\o$.

To summarize, the formalism of broken replica
 symmetry  provides a large space for the probability distribution
of $P(\o)$ (when one changes sample). The probability distributions
so generated
 have the following properties:

 a) They depend on many parameters so that we have a large variety of
 choices.

 b) If we want, we can construct these probabilities in
  such a way that the system is scaling
 invariant at large distances, characterized for instance by a very
non trivial scaling exponent $\zeta$.

 c) The expectation values can be computed explicitly so
 that this solution
can be taken as the starting point of a perturbative expansion.

 d) Last, but not least, the gaussian rsb Ansatz
   becomes exact when the dimension of the space
 goes to infinity.

An interesting example of application of this very general
scheme to the vortices in superconductors
can be found in \cite{boumezyed,gialed,korshu,gialedrev}.

\acknowledgements
It is a great pleasure to thank G. Parisi
with whom many of the ideas presented here have been
developed.

\end{document}